\documentclass[12pt]{article}
\usepackage[english]{babel}
\usepackage{latexsym}
\usepackage{amsfonts}
\begin{document}
\rightline{NPI MSU 2002-18/702}
\rightline{July 2002}
\vspace{1cm}
\begin{center}
{\Large \bf Linearized gravity, Newtonian limit\\ and light
deflection in RS1 model}\\

\vspace{4mm}

Mikhail N.~Smolyakov$^b$, Igor P.~Volobuev$^a$\\

\vspace{4mm}

$^a$ Skobeltsyn Institute of Nuclear Physics, Moscow State
University
\\ 119899 Moscow, Russia \\
$^b$ Physical Department, Moscow State University \\ 119899
Moscow, Russia \\

\end{center}

\begin{abstract}
We solve exactly the equations of motion for linearized gravity in
the Randall-Sundrum model with matter on the branes and calculate
the Newtonian limit in it. The result contains contributions of
the radion and  of the massive modes, which change considerably
Newton's law at small distances. The effects of "shadow"\ matter,
which lives on the other brane, are considered and compared with
those of ordinary matter for both positive and negative tension
branes. We also calculate light deflection and Newton's law in the
zero mode approximation and explicitly distinguish the
contribution of the radion field. \\[0.3cm] Keywords: Kaluza-Klein
theories; branes;  linearized gravity; Newtonian limit
\end{abstract}

\section{Introduction}

The present-day approach to Kaluza-Klein  theories is based on the
idea due to Rubakov and Shaposhnikov of localization of fields on
a domain wall \cite{RSH,RSH1}. In a first step it is natural to
drop the mechanism of localization of fields and to treat the
domain wall as an infinitely thin object, i.e. as a membrane, and
to consider the effects due to the gravitational interaction of
such 3-branes.

A particular realization of this scenario was proposed in paper \cite{RS1}.
In this paper an exact solution for a system of two branes interacting with
gravity in a five-dimensional space-time was found. This model is called
the Randall-Sundrum model (usually abbreviated as RS1 model), and it is
widely discussed in the literature (see Refs. \cite{Rub01,Ant01} for
reviews and references). A consistent analysis of this model without matter
on the branes was made in \cite{BKSV}.  However, the equations of motion
for metric fluctuations, when there is matter located on the branes, have
not been studied in detail yet. In paper \cite{GarTan} the Randall-Sundrum
model with matter on the branes was discussed for the cases of both one and
two branes (the former is usually called RS2 model). This paper utilized
the Gaussian normal coordinates and the bent-brane formalism, which was
argued to be inconsistent in RS2 model in paper \cite{AIMVV}. Moreover, the
use of Gaussian normal coordinates mixed the contributions of the graviton
and the radion fields to the four-dimensional gravitational field. In the
present paper we will decouple the equations for the graviton and the
radion fields and solve exactly the equations of motion for the RS1 model
with matter on the branes. Then we will calculate the Newtonian limit in
this model and the light deflection by a point-like mass in the zero mode
approximation.

The Randall-Sundrum model \cite{RS1} describes the gravity in a
five-dimensional space-time $E$ with two branes embedded into it. We
denote the coordinates by $ \{ x^M\} \equiv \{x^{\mu},y\}$,
$M= 0,1,2,3,4, \, \mu=0,1,2,3$, the coordinate $x^4 \equiv y$  parameterizing
the fifth dimension. It forms the orbifold $S^{1}/Z_{2}$, which is
realised as the circle of the circumference $2R$ with points $y$
and $-y$ identified. Correspondingly, we have the usual periodicity
condition in space-time $E$, which identifies points $(x, y)$ and
$(x, y + 2nR)$, and
the metric $g_{MN}$ satisfies the orbifold symmetry conditions
\begin{eqnarray}
\label{orbifoldsym}
 g_{\mu \nu}(x,- y)=  g_{\mu \nu}(x,  y), \\
 \nonumber
  g_{\mu 4}(x,- y)= - g_{\mu 4}(x,  y), \\ \nonumber
   g_{44}(x,- y)=  g_{44}(x,  y).
\end{eqnarray}
The branes are located at the fixed points of the orbifold, $y=0$
and $y=R$.

The action of the model is
\begin{equation}\label{actionRS}
 S = S_g + S_1 + S_2,
\end{equation}
where $S_g$, $S_1$ and $S_2$ are given by
\begin{eqnarray}\label{actionsRS}
S_g&=& \frac{1}{16 \pi \hat G} \int_E
\left(R-\Lambda\right)\sqrt{-g}\, d^{4}x dy,\\ \nonumber
 S_1&=& V_1 \int_E \sqrt{-\tilde g} \delta(y) d^{4}x dy,\\ \nonumber
 S_2&=& V_2 \int_E \sqrt{-\tilde g}  \delta(y-R) d^{4}x dy.
\end{eqnarray}
{Here $\tilde g_{\mu\nu}$ is the induced metric on the branes and the
subscripts 1 and 2 label the branes.} We also note that the signature of
the metric $g_{MN}$ is chosen to be $(-1,1,1,1,1)$.

The Randall-Sundrum solution for the  metric is {given by}
\begin{equation}\label{metricrs}
ds^2=  g_{MN} d{x}^M d{x}^N = e^{2\sigma(y)} \eta_{\mu\nu}
{dx^\mu dx^\nu} +
  dy^2,
\end{equation}
where $\eta_{\mu\nu}$ is the Minkowski metric and  {the function}
$\sigma(y) = -k|y|$ in the interval $-R \leq y \leq R$. The
parameter  $k$ is positive and  has the dimension of mass, the
parameters $\Lambda$ and $ V_{1,2}$ are {related to it as
follows:} $$ \Lambda = -12 k^2, \quad V_1 =-V_2= -\frac{3k}{4\pi
\hat G}. $$ We see that brane~1 has a positive energy density,
whereas brane~2 has a negative one. The function $\sigma$ has the
properties
\begin{equation}\label{sigma}
  \partial_4 \sigma = -k\, sign(y), \quad \frac{\partial^2 \sigma}{\partial
  {y}^2} =-2k(\delta(y) - \delta(y-R)) \equiv  -2k\tilde \delta .
\end{equation}
Here and in the sequel $\partial_4 \equiv \frac{\partial}{\partial y}$.

We denote $\hat \kappa = \sqrt{16 \pi \hat G}$, where $\hat G$ is the
five-dimensional gravitational constant, and parameterize the metric
$g_{MN}$ as
\begin{equation}\label{metricpar}
  g_{MN} = \gamma_{MN} + \hat \kappa h_{MN},
\end{equation}
$h_{MN}$ being the metric fluctuations. Substituting this parameterization
into (\ref{actionRS}) and retaining the terms of the zeroth order in $\hat \kappa$,
we get the second variation action of this model  \cite{BKSV}. It  is invariant
under the gauge transformations
\begin{eqnarray}\label{gaugetrRS}
h'_{MN}(x,y) = h_{MN}(x,y) -(\nabla_M\xi_N(x,y) + \nabla_N\xi_M(x,y) ),
\end{eqnarray}
where $\nabla_M$ is the covariant derivative with respect to the
background metric $\gamma_{MN}$, and the functions $\xi_N(x,y)$
satisfy the orbifold symmetry conditions
\begin{eqnarray}\label{orbifoldsym1}
\xi^{\mu}\left(x,-y\right)&=&\xi^{\mu}\left(x,y\right),\\
\nonumber \xi^{4}\left(x,-y\right)&=&-\xi^{4}\left(x,y\right).
\nonumber
\end{eqnarray}
With the help of these gauge transformations we can impose the
gauge
\begin{equation}\label{unitgauge}
h_{\mu4} =0, \, h_{44} = h_{44}(x) \equiv \phi (x),
\end{equation}
which  will be called the {\it unitary gauge} (see \cite{BKSV}). We would
like to emphasize once again that the branes remain straight in this gauge,
i.e. we {\it do not} use the bent-brane formulation, which allegedly
destroys the structure of the model (this problem was discussed in
\cite{AIMVV}).

The general form of interaction with matter is standard,
\begin{equation}\label{interaction}
 \frac{\hat \kappa}{2} \int_{B_1} h^{\mu\nu}(x,0) T^1_{\mu\nu} dx +
  \frac{\hat \kappa}{2} \int_{B_2} h^{\mu\nu}(x,R) T^2_{\mu\nu} \sqrt{- \gamma_2} dx,
\end{equation}
where $ T^1_{\mu\nu}$ and $ T^2_{\mu\nu}$ are energy-momentum
tensors of the matter on brane~1 and brane~2 respectively:
$$
 T^{1,2}_{\mu\nu} = 2\frac{\delta L^{1,2}}{\delta  \gamma^{\mu\nu}} -
  \gamma^{1,2}_{\mu\nu}  L^{1,2}.
$$

As follows from formula (\ref{interaction}), $h_{\mu\nu}$ is the only
physical field of the model, since only this field interacts with  matter
on the branes. Obviously, the unitary gauge conditions (\ref{unitgauge}) do
not fix the gauge of this field. In fact, after imposing these gauge
conditions there remain gauge transformations of the form
\begin{equation}\label{remgaugetr}
   \xi_\mu = e^{2\sigma}\epsilon_\mu(x),
\end{equation}
which change the longitudinal components of the field $h_{\mu\nu}$.
Nevertheless, it turns out that it is convenient to  solve the equations of
motion for linearized gravity in the unitary gauge and then to  choose an
appropriate gauge  in our four-dimensional world on the brane. We will use
the de Donder gauge for the field $h_{\mu\nu}$  on the branes, which
corresponds  to the choice of  {\it harmonic coordinates}.

\section{Newtonian limit}
The equations of motion  for different components of the metric
fluctuations in the unitary gauge take the form (see \cite{BKSV}):

 1) $\mu\nu$-component
\begin{eqnarray}\label{mu-nu}
 & &\frac{1}{2}\left(\partial_\rho \partial^\rho h_{\mu\nu}-
\partial_\mu \partial^\rho
h_{\rho\nu}-\partial_\nu \partial^\rho h_{\rho\mu} +
\frac{\partial^2 h_{\mu\nu}}{\partial {x^4}^2}\right)- \\ \nonumber
&-& 2k^2h_{\mu\nu}+\frac{1}{2}\partial_\mu \partial_\nu\tilde h+
\frac{1}{2}\partial_\mu \partial_\nu \phi+  \\ \nonumber
&+& \frac{1}{2} \gamma_{\mu\nu}\left(\partial^\rho \partial^\sigma
h_{\rho\sigma}-\partial_\rho \partial^\rho \tilde h -
\frac{\partial^2 \tilde h}{\partial {x^4}^2}-4\partial_4 \sigma
\partial_4 \tilde h
 - \partial_\rho \partial^\rho \phi + 12 k^2 \phi\right)+\\ \nonumber
&+& \left[2k  h_{\mu\nu} - 3k\gamma_{\mu\nu}\phi \right]\tilde
\delta = -\frac{\hat \kappa}{2} T_{\mu\nu},
\end{eqnarray}

 2) $\mu 4$-component,
\begin{equation}\label{mu-4}
\partial_4 ( \partial_\mu \tilde h - \partial^\nu  h_{\mu\nu})-
3\partial_4 \sigma \partial_\mu \phi = 0,
\end{equation}
which plays the role of a constraint,

 3) $4 4$-component
\begin{equation}\label{4-4}
\frac{1}{2}(\partial^\mu \partial^\nu  h_{\mu\nu} - \partial_\mu
\partial^\mu \tilde h ) - \frac{3}{2}\partial_4 \sigma \partial_4 \tilde h
+ 6 k^2 \phi =0,
\end{equation}
with $T_{\mu\nu}$ being the energy-momentum tensor of the matter
and $\tilde h=\gamma^{\mu\nu}h_{\mu\nu}$. In what follows, we will
also use an auxiliary equation, which is obtained by multiplying
the equation for $44$-component by 2 and subtracting it from the
contracted equation for $\mu\nu$-component. This equation contains
$\tilde h$ and $\phi$ only and has the form:
\begin{equation}\label{contracted-44}
\frac{\partial^2 \tilde h}{\partial {x^4}^2}  + 2\partial_4 \sigma \partial_4 \tilde h
  -8k^2 \phi+ 8k \phi \tilde \delta + \partial_\mu \partial^\mu \phi =
\frac{\hat \kappa}{3} T_{\mu}^{\mu}.
\end{equation}

If $ T_{\mu\nu}=0,$ the physical degrees of freedom of  the model
can be extracted by the substitution \cite{BKSV}
\begin{equation}\label{substitution}
 h_{\mu\nu} =  b_{\mu\nu} + \gamma_{\mu\nu}(\sigma - c)\phi +
 \frac{1}{2k^2} \left(\sigma - c +\frac{1}{2} +
 \frac{c}{2}e^{-2\sigma}\right) \partial_\mu \partial_\nu \phi.
\end{equation}
with $c=\frac{kR}{e^{2kR}-1}$. It turns out that the field
$b_{\mu\nu}(x^{\mu},y)$ describes the massless graviton
\cite{RS1,RS2} and massive Kaluza-Klein spin-2 fields, whereas
$\phi (x)$ describes a scalar field called the radion.
Apparentely, the radion field was first identified in Ref.
\cite{ADM} (see also \cite{Sundr}) and discussed
in  \cite{CGR,CGRT,wise2,Das}.

However, the situation is rather different, when there is matter on the
branes. Let us first  consider the case, where matter is located on the
brane at $0$, i.e. the energy-momentum tensor is of the form $T_{\mu\nu} =
 t_{\mu\nu}(x)\delta(y)$. The substitution, which allows one to decouple
the equations, looks like
\begin{eqnarray}\label{substitution1}
h_{\mu\nu} =  u_{\mu\nu} + \gamma_{\mu\nu}(\sigma - c)\phi +
 \frac{1}{2k^2} \left(\sigma - c +\frac{1}{2} +
{c}e^{2kR}\right) \partial_\mu \partial_\nu \phi.
\end{eqnarray}
After this substitution equations (\ref{mu-4}), (\ref{4-4}) and
(\ref{contracted-44}), rewritten in the flat metric (i.e. $u=
\eta^{\mu\nu}u_{\mu\nu}$), take the form:
\begin{eqnarray}\label{mu4TT}
\partial_4(e^{-2\sigma}(\partial^\nu u_{\mu\nu}-\partial_\mu
u))=0,
\end{eqnarray}
\begin{eqnarray}\label{44TT}
e^{-4\sigma}(\partial^\mu\partial^\nu u_{\mu\nu}-\Box u) -
3\partial_4\sigma\partial_4(e^{-2\sigma}u)+3ce^{2kR-2\sigma}\Box
\phi=0,
\end{eqnarray}
\begin{eqnarray}\label{contTT}
\partial_4(e^{2\sigma}\partial_4(e^{-2\sigma}u))+2\frac{c}{k}\delta(y)\left[e^{2kR}
-1\right]\Box \phi=\frac{\hat\kappa}{3}t_\mu^\mu \, \delta(y).
\end{eqnarray}

Let us consider Fourier expansion of all terms of  equation
(\ref{contTT}) with respect to coordinate $y$. Since the term
with the derivative $\partial_4$ has no zero mode, this equation implies that
\begin{equation}\label{eqphi}
\Box \phi=\frac{\hat\kappa}{6R}t, \quad t \equiv t_\mu^\mu ,
\end{equation}
\begin{equation}\label{equ}
\partial_4(e^{-2\sigma}u)=0.
\end{equation}
From the last equation and equation (\ref{mu4TT}) it follows that
\begin{eqnarray}
\partial^{\nu}u_{\mu\nu}=e^{2\sigma}A_{\mu}(x), \\
\partial_{\mu}u^{\nu}_{\nu}=e^{2\sigma}B_{\mu}(x),
\end{eqnarray}
where $A_{\mu}(x)$ and $B_{\mu}(x)$ depend on four-dimensional coordinates
only. It is easy to see that the remaining  gauge transformations
(\ref{remgaugetr}) allow us to impose the de Donder gauge condition on the
field $u_{\mu\nu}$
\begin{equation}\label{dedonder}
\partial^\nu\left(u_{\mu\nu}-\frac{1}{2}\eta_{\mu\nu}
u\right)=0.
\end{equation}
Having imposed this gauge, we are still left with residual gauge
transformation
\begin{equation}\label{ostatgauge}
\xi_\mu = e^{2\sigma}\epsilon_\mu(x), \quad \Box\epsilon_{\mu}=0.
\end{equation}
The gauge transformations with $\xi_\mu$ satisfying these
conditions are important for determining the number of degrees of
freedom of the massless graviton. It follows from equation
(\ref{44TT}) that
\begin{equation}\label{boxu}
\Box u=e^{2\sigma}\frac{\hat\kappa k}{1-e^{-2kR}}t.
\end{equation}
Now let us consider $\mu\nu$-equation. After substitution
(\ref{substitution1}) and in the de Donder gauge it takes the form
\begin{eqnarray}\label{munuDeD}
\frac{1}{2}e^{-2\sigma}\Box u_{\mu\nu}+\frac{1}{2}\partial_4
\partial_4 u_{\mu\nu}-2k^2u_{\mu\nu}-\partial_4
\partial_4\sigma u_{\mu\nu}= \\ \nonumber
=-\frac{\hat\kappa}{2}f_{\mu\nu}\delta(y)+\frac{\hat\kappa
k}{12(1-e^{-2kR})}\left(\eta_{\mu\nu}+2\frac{\partial_\mu\partial_\nu}{\Box}
\right)t,
\end{eqnarray}
where
\begin{equation}
f_{\mu\nu}=t_{\mu\nu}-\frac{1}{3}\left(\eta_{\mu\nu}-
\frac{\partial_\mu\partial_\nu}{\Box}\right)t.
\end{equation}
One can see that $f_{\mu\nu}$ is transverse-traceless. Here and below
the  inverse d'Alembertian is an integral operator uniquely  defined by
the  radiation conditions.

To solve this equation, we make the following substitution
\begin{equation}\label{sub1}
u_{\mu\nu}=v_{\mu\nu}+\frac{\hat\kappa
k}{6(1-e^{-2kR})}e^{2\sigma}\Box^{-1}\left(\eta_{\mu\nu}+
2\frac{\partial_\mu\partial_\nu}{\Box}\right)t,
\end{equation}
which does not violate (\ref{dedonder}) that takes the form
\begin{eqnarray}\label{dedonder1}
\partial^\nu v_{\mu\nu}=0, \\ \nonumber
v_\mu^\mu=0.
\end{eqnarray}

Substituting (\ref{sub1}) into (\ref{munuDeD}), we get
\begin{eqnarray}\label{munuDeD1}
\frac{1}{2}e^{-2\sigma}\Box v_{\mu\nu}+\frac{1}{2} \frac{\partial^2}{
\partial y^2} v_{\mu\nu}-2k^2v_{\mu\nu}- \frac{\partial^2}{ \partial
y^2}\sigma v_{\mu\nu}= -\frac{\hat\kappa}{2}f_{\mu\nu}\delta(y).
\end{eqnarray}
This equation can be solved  exactly. To  this end, let us first
perform Fourier transform of the field $v_{\mu\nu}$ with respect
to the $x$-coordinates
\begin{eqnarray}\label{munuDeD2}
\frac{1}{2}e^{-2\sigma}(-p^2) \tilde
v_{\mu\nu}+\frac{1}{2}\frac{\partial^2}{ \partial y^2}\tilde
v_{\mu\nu}-2k^2\tilde v_{\mu\nu}-\frac{\partial^2}{ \partial y^2}\sigma
\tilde v_{\mu\nu}=-\frac{\hat\kappa}{2}\tilde f_{\mu\nu}\delta(y),
\end{eqnarray}
where $p^2=-p_{0}^{2}+\bar p^2$.

First let us solve this equation in the bulk (see \cite{BKSV},
\cite{RS2}). Here the solution of equation (\ref{munuDeD2})
looks like
\begin{eqnarray}
\tilde
v_{\mu\nu}(p,y)=C_{\mu\nu}J_2\left(\frac{\sqrt{|p^2|}}{k}e^{k|y|}\right)+
D_{\mu\nu}N_2\left(\frac{\sqrt{|p^2|}}{k}e^{k|y|}\right), p^2<0 \\
\tilde
v_{\mu\nu}(p,y)=C_{\mu\nu}I_2\left(\frac{\sqrt{|p^2|}}{k}e^{k|y|}\right)+
D_{\mu\nu}K_2\left(\frac{\sqrt{|p^2|}}{k}e^{k|y|}\right), p^2>0
\end{eqnarray}
Substituting it into  equation (\ref{munuDeD2}) and comparing the
terms at the boundaries, we get the values of the constant tensors
$C_{\mu\nu}$ and $D_{\mu\nu}$. Having got them, we obtain the
solutions of the $\mu\nu$-equation with matter on the
positive tension brane:

1)$p^2<0$
\begin{eqnarray}
\tilde v_{\mu\nu}(p,y)= \left[\tilde
t_{\mu\nu}-\frac{1}{3}\left(\eta_{\mu\nu}-
\frac{p_{\mu}p_{\nu}}{p^2}\right)\tilde t \right]
\frac{\hat\kappa}{2\sqrt{-p^2}}\times \\ \nonumber \times
\frac{N_2\left(\frac{\sqrt{-p^2}}{k}e^{k|y|}\right)
J_1\left(\frac{\sqrt{-p^2}}{k}e^{kR}\right)-J_2\left(\frac{\sqrt{-p^2}}{k}e^{k|y|}\right)
N_1\left(\frac{\sqrt{-p^2}}{k}e^{kR}\right)}{J_1\left(\frac{\sqrt{-p^2}}{k}\right)
N_1\left(\frac{\sqrt{-p^2}}{k}e^{kR}\right)-N_1\left(\frac{\sqrt{-p^2}}{k}\right)
J_1\left(\frac{\sqrt{-p^2}}{k}e^{kR}\right)},
\end{eqnarray}

2)$p^2>0$
\begin{eqnarray}\label{pbigger0}
\tilde v_{\mu\nu}(p,y)=-\left[\tilde
t_{\mu\nu}-\frac{1}{3}\left(\eta_{\mu\nu}-\frac{p_{\mu}p_{\nu}}{p^2}\right)\tilde
t \right]\frac{\hat\kappa}{2\sqrt{p^2}}\times \\ \nonumber
\times\frac{K_2\left(\frac{\sqrt{p^2}}{k}e^{k|y|}\right)
I_1\left(\frac{\sqrt{p^2}}{k}e^{kR}\right)+I_2\left(\frac{\sqrt{p^2}}{k}e^{k|y|}\right)
K_1\left(\frac{\sqrt{p^2}}{k}e^{kR}\right)}{I_1\left(\frac{\sqrt{p^2}}{k}\right)
K_1\left(\frac{\sqrt{p^2}}{k}e^{kR}\right)-I_1\left(\frac{\sqrt{p^2}}{k}e^{kR}\right)
K_1\left(\frac{\sqrt{p^2}}{k}\right)}.
\end{eqnarray}
Since we want to calculate the Newtonian limit, $\tilde t_{\mu\nu}$
is proportional to $\delta(p_0)$ in this case. It means, that we
need the solution for $p^2>0$.

When  matter is  on brane~2 (at $y=R$), all the reasonings are the
same, as presented above. The full substitution looks like
\begin{eqnarray}\label{subsT2-1}
h_{\mu\nu} = u_{\mu\nu}+e^{2\sigma}\eta_{\mu\nu}(\sigma - c)\phi +
 \frac{1}{2k^2} \left(\sigma  +\frac{1}{2}\right)
 \partial_\mu \partial_\nu \phi, \\ \nonumber
u_{\mu\nu}=v_{\mu\nu} + \frac{\hat\kappa
k}{6(e^{2kR}-1)}e^{2\sigma}\Box^{-1}\left(\eta_{\mu\nu}+
2\frac{\partial_\mu\partial_\nu}{\Box}\right)t.
\end{eqnarray}
The gauge conditions are the same too. The equations of motion for
the field $\phi$ and solution for the field $v_{\mu\nu}$ with matter
on brane~2 can be derived  analogously and read
\begin{equation}\label{eqphi2}
\Box \phi=\frac{\hat\kappa}{6R}t,
\end{equation}

1)$p^2<0$
\begin{eqnarray}
\tilde v_{\mu\nu}(p,y)=\left[\tilde
t_{\mu\nu}-\frac{1}{3}\left(\eta_{\mu\nu}-\frac{p_{\mu}p_{\nu}}{p^2}\right)\tilde
t \right]\frac{\hat\kappa}{2\sqrt{-p^2}}\frac{1}{e^{kR}}\times \\ \nonumber
\times \frac{N_2\left(\frac{\sqrt{-p^2}}{k}e^{k|y|}\right)
J_1\left(\frac{\sqrt{-p^2}}{k}\right)-J_2\left(\frac{\sqrt{-p^2}}{k}e^{k|y|}\right)
N_1\left(\frac{\sqrt{-p^2}}{k}\right)}{J_1\left(\frac{\sqrt{-p^2}}{k}\right)
N_1\left(\frac{\sqrt{-p^2}}{k}e^{kR}\right)-N_1\left(\frac{\sqrt{-p^2}}{k}\right)
J_1\left(\frac{\sqrt{-p^2}}{k}e^{kR}\right)} ,
\end{eqnarray}

2)$p^2>0$
\begin{eqnarray}
\tilde v_{\mu\nu}(p,y)=-\left[\tilde
t_{\mu\nu}-\frac{1}{3}\left(\eta_{\mu\nu}-\frac{p_{\mu}p_{\nu}}{p^2}\right)\tilde
t \right]\frac{\hat\kappa}{2\sqrt{p^2}}\frac{1}{e^{kR}}\times \\ \nonumber
\times \frac{K_2\left(\frac{\sqrt{p^2}}{k}e^{k|y|}\right)
I_1\left(\frac{\sqrt{p^2}}{k}\right)+I_2\left(\frac{\sqrt{p^2}}{k}e^{k|y|}\right)
K_1\left(\frac{\sqrt{p^2}}{k}\right)}{I_1\left(\frac{\sqrt{p^2}}{k}\right)
K_1\left(\frac{\sqrt{p^2}}{k}e^{kR}\right)-I_1\left(\frac{\sqrt{p^2}}{k}e^{kR}\right)
K_1\left(\frac{\sqrt{p^2}}{k}\right)}.
\end{eqnarray}
An important point is that these equations are written in the
coordinates $\{x^{\mu}\}$, which are {\it Galilean on brane~1}
(not on brane~2) and are inappropriate for studying physical
effects on brane~2 (we recall that coordinates are called Galilean,
if $g_{\mu \nu} = diag(-1, 1,1,1)$ \cite{LL}).

Obviously, when there is  matter on both branes, the solution for
the  metric fluctuation  is just the sum of solutions for each
brane separately, which follows from the  linearity of the
equations.

Now let us examine the four-dimensional theory on the branes. First we
consider the case when matter and observer are located on brane~1.
It is easy to see that $h_{\mu\nu}$ (\ref{substitution1}),
(\ref{sub1}) does not satisfy the de Donder gauge condition. The
residual gauge transformations (\ref{ostatgauge}) are not
sufficient to pass to this gauge. But since we consider only the
effective theory on brane~1, we can drop the
$\Box\epsilon_{\mu}=0$ condition, which fixes the gauge for the
field $v_{\mu\nu}$. Then we can pass to the  de Donder gauge
condition for the field  $h_{\mu\nu}$ {\it on brane~1} with the
help of the gauge functions of the form
\begin{equation}\label{gfunc}
\epsilon_{\mu}=\frac{1}{2}\left(\frac{R}{2k}+\frac{1}{4k^2}\right)\partial_{\mu}\phi+
\frac{kR}{e^{2kR}-1}\frac{\partial_{\mu}}{\Box} \ \phi.
\end{equation}
Having been made, these transformations result in
\begin{equation}\label{honbrane1}
\left. h_{\mu\nu}\right|_{y=0}=\left.
v_{\mu\nu}\right|_{y=0}+\eta_{\mu\nu}kR\phi+
2kR\frac{\partial_{\mu}\partial_{\nu}}{\Box}\phi.
\end{equation}
This formula gives the four-dimensional gravitational
field on brane~1, and it takes into account the  contributions of the massless and
of the massive gravitons and the contribution of the radion.

Now let us calculate the Newtonian limit in this case. Let us consider
a static point mass with energy-momentum tensor
$t_{00}=M\delta(\vec x), t_{0k}=t_{ij}=0$ (and $\tilde t_{00}=2\pi
M \delta (p_0) $), $M$ denoting the inertial mass. We need to
calculate only $h_{00}$-component of the metric fluctuations (see
(\ref{pbigger0}))
\begin{eqnarray}\label{pbigger0newt}
\tilde
v_{00}(p,0)=-\frac{\hat\kappa}{2\sqrt{p^2}}\frac{K_2\left(\frac{\sqrt{p^2}}{k}\right)
I_1\left(\frac{\sqrt{p^2}}{k}e^{kR}\right)+I_2\left(\frac{\sqrt{p^2}}{k}\right)
K_1\left(\frac{\sqrt{p^2}}{k}e^{kR}\right)}{I_1\left(\frac{\sqrt{p^2}}{k}\right)
K_1\left(\frac{\sqrt{p^2}}{k}e^{kR}\right)-I_1\left(\frac{\sqrt{p^2}}{k}e^{kR}\right)
K_1\left(\frac{\sqrt{p^2}}{k}\right)}\times \\ \nonumber
\times\frac{4\pi}{3}M\delta (p_0)
\end{eqnarray}
and
\begin{eqnarray}
h_{00}(x,0)=v_{00}(x,0)-kR\frac{1}{4\pi}\frac{\hat\kappa}{6R}\frac{M}{r},
\end{eqnarray}
(see (\ref{honbrane1})) where $r=\sqrt{{\vec x}^2}$ and
\begin{equation}
v_{00}(x,0)=\frac{1}{(2\pi)^4}\int e^{-i\eta_{\mu\nu}p^\mu
x^\nu}\tilde v_{00}(p,0)d^4p.
\end{equation}
Using the last relation and equation (\ref{pbigger0newt}) one can
easy find that
\begin{eqnarray}\label{integral}
v_{00}(x,0)&=&\frac{1}{{(2\pi)}^4}\frac{4\pi}{{r}}\frac{4\pi}{3}M\int_0^\infty
g(p)sin(pr)p dp,\\
g(p)&=&\frac{\hat\kappa}{2p}\frac{K_2\left(\frac{p}{k}\right)
I_1\left(\frac{p}{k}e^{kR}\right)+I_2\left(\frac{p}{k}\right)
K_1\left(\frac{p}{k}e^{kR}\right)}{I_1\left(\frac{p}{k}\right)
K_1\left(\frac{p}{k}e^{kR}\right)-I_1\left(\frac{p}{k}e^{kR}\right)
K_1\left(\frac{p}{k}\right)}, \nonumber
\end{eqnarray}
where $p=\sqrt{{\vec p}^2}$. It is impossible to evaluate the integral
in (\ref{integral}) analytically. But we can estimate the integrand in the
following cases: 1) $p<<e^{-kR}k$, 2) $e^{-kR}k<<p<<k$ and 3) $k<<p$. Using
these estimates in the intervals $0\le p\le e^{-kR}k$, $e^{-kR}k\le p\le k$
and $k<p<\infty$ respectively, we can estimate the integral in equation
(\ref{integral}).

First, utilizing the recurrent relations for the McDonald functions, we
represent the  factor $g(p)$ in the integrand in the following form
\begin{equation}\label{antrans}
g(p) =\frac{\hat\kappa
k}{p^2}-\frac{\hat\kappa}{2p}\frac{K_0\left(\frac{p}{k}\right)
I_1\left(\frac{p}{k}e^{kR}\right)+I_0\left(\frac{p}{k}\right)
K_1\left(\frac{p}{k}e^{kR}\right)}{I_1\left(\frac{p}{k}\right)
K_1\left(\frac{p}{k}e^{kR}\right)-I_1\left(\frac{p}{k}e^{kR}\right)
K_1\left(\frac{p}{k}\right)}.
\end{equation}
With the help of this transformation we have picked out the contribution
of the zero mode to $v_{00}$. Substituting asymptotic formulas for McDonald
functions into (\ref{antrans}), we get
\begin{eqnarray}\label{estimates}
g(p)\approx\frac{\hat\kappa k}{p^2}+{\left[
 \begin{array}{lcl}
\frac{\hat\kappa k}{e^{2kR}-1}\frac{1}{p^2}-
\frac{\hat\kappa}{2k\left(1-e^{-2kR}\right)}\,
ln\left(\frac{p}{k}\right)&, &p<<e^{-kR}k \\
-\frac{\hat\kappa}{2k}\, ln\left(\frac{p}{k}\right)&,
&e^{-kR}k<<p<<k \\ \frac{\hat\kappa}{2p}&,  &k<<p. \\
\end{array}
\right.}
\end{eqnarray}
All the integrals arising after substituting (\ref{estimates}) into
(\ref{integral}) can be calculated analytically. The integrals of the
form $\int_a^\infty sin(pr) dp$ are calculated with the use of the
regularization
\begin{equation}
\int_a^\infty sin(pr) dp=\lim_{\epsilon\to 0
}\left(\int_{a}^{\infty} e^{-\epsilon p}\sin(pr) dp
\right)=\frac{\cos(ar)}{r}.
\end{equation}
It is well known that the gravitational potential is expressed
through the $g_{00}$-component of the  metric as
$g_{00}=-1-2V$ \cite{Weinberg}. Since $g_{00}=-1+\hat\kappa
h_{00}$, we get
\begin{equation}
V=-\frac{\hat\kappa}{2}h_{00}.
\end{equation}
Thus, using equations
(\ref{honbrane1}), (\ref{integral}) and (\ref{estimates}), we get
\begin{equation}\label{NL1}
V\approx -\frac{M}{r}G_1\left(1+\frac{4\cos(kr)}{3\pi
kr}-\frac{4}{3\pi k^2 r^2}\left[sin(kr)-si(kr)\right]\right),
\end{equation}
where $G_1=\frac{\hat G k}{1-e^{-2kR}}$ and $$si(b)=\int_0^b
\frac{\sin{t}}{t}\, dt.$$

Thus, we have examined the case of the mass and the observer being
located on brane~1. But  there are three more  possible cases to
be examined. It is the case  of "shadow"\ matter, when the mass is
located on brane~2 and the observer is located on brane~1, the
case of the mass and the observer being  located on brane~2, and
the case of the  mass being located on brane~1, whereas the
observer being located on brane~2 ("shadow"\ matter effect as
well). All the calculations in these cases are the same, as
described above. But if the  observer is located on brane~2, it is
necessary to pass to Galilean coordinates on brane~2 to get a
correct result. This problem was discussed in detail in paper
\cite{BKSV}. The energy-momentum tensor $t_{\mu\nu}$ takes the
form $t_{00}=M\delta (\vec x ), t_{0k}=t_{ij}=0$ in the Galilean
coordinates on the brane the mass is located on. In the formulas
presented below (in the Galilean coordinates on the observer's
brane) the energy-momentum tensor always has this canonical form,
and passing to the Galilean coordinates is taken into account by a
factor in front of the brackets.

Now let us discuss, how to pass to Galilean coordinates on brane~2
by the example of the case when the mass is located on brane~1 and
the observer is located on brane~2 ("shadow"\ matter effect). We
consider the following form of the Fourier transform with the
integrand   appropriate for this case
\begin{equation}
A_{\mu\nu}(x_0,\vec x)=\frac{1}{(2\pi)^4} B_{\mu\nu} \,\int
e^{i\eta_{\mu\nu}p^{\mu}x^{\nu}}g(p)\delta(p_0) d^{4}p,
\end{equation}
with $B_{\mu\nu}$ being a constant. One can easily find, that
in Galilean coordinates on brane~2 this expression looks like
\begin{equation}
A'_{\mu\nu}(z_0,\vec z)=\frac{1}{(2\pi)^4} B_{\mu\nu}\, e^{-kR}\int
e^{i\eta_{\mu\nu}q^{\mu}z^{\nu}}g(e^{-kR}q)\delta(q_0) d^{4}q,
\end{equation}
where $\{z^{\mu}\}$ are Galilean coordinates on brane~2. In particular,
this means, that the intervals, in which we have to estimate the
integrand, change. Now we have to choose the following domains of
estimate: $q<<k$, $k<<q<<e^{kR}k$ and $e^{kR}k<<q$.

Another difficulty, which arises in this case, is that one of the integrals
cannot be evaluated analytically. It has the following form
\begin{equation}\label{comp}
\int_k^\infty
e^{-\frac{q}{k}}\sqrt{q}\sin(qr)\,dq=k^{3/2}\int_1^\infty
\sqrt{t}\sin(tkr)e^{-t}\,dt,
\end{equation}
where $r=\sqrt{{\vec z}^2}$. But we can estimate it in the
limiting cases $1<<kr$ and $kr<<1$.

1) $1<<kr$. In this case (\ref{comp}) can be estimated by
integration by parts
\begin{equation}
\int_1^\infty \sqrt{t}\sin(tkr)e^{-t}\,dt\approx
e^{-1}\frac{\cos(kr)}{kr}.
\end{equation}

2) $kr<<1$. In this case we can make the expansion of\, $sin(tkr)$ for
the small argument
\begin{eqnarray}
\int_1^\infty \sqrt{t}\sin(tkr)e^{-t}\,dt\approx \int_0^\infty
\sqrt{t}\sin(tkr)e^{-t}\,dt- \\ \nonumber
-\int_0^1 \sqrt{t}\left(krt-\frac{(kr)^3t^3}{6}+...\right)e^{-t}\,dt.
\end{eqnarray}
The first integral was taken from the table of integrals
\cite{Integ} and the second one was calculated numerically. The
result reads as follows
\begin{equation}
\int_1^\infty \sqrt{t}\sin(tkr)e^{-t}\,dt\approx \
\left(\frac{3}{4}\sqrt{\pi}-0.2\right)kr+0.099\frac{1}{6}\,(kr)^3.
\end{equation}

And the last very interesting point of "shadow"\ matter cases is
passing to the de Donder gauge. Let us show it in the case of the
field $\left. h_{\mu\nu}\right|_{y=R}$. It does not satisfy the de
Donder gauge condition on brane~2 (see (\ref{subsT2-1})),
analogously to the case of matter located on brane~1. With the
help of the gauge functions of the following form
\begin{equation}
\epsilon_\mu=\frac{e^{2kR}}{8k^2}\partial_{\mu}\phi+
\frac{kRe^{2kR}}{e^{2kR}-1}\frac{\partial_{\mu}}{\Box}\phi
\end{equation}
we can pass to the de Donder gauge. But after this transformation
$\left. h_{\mu\nu}\right|_{y=R}$ appears to have the following
form
\begin{equation}
\left. h_{\mu\nu}\right|_{y=R}=\left. v_{\mu\nu}\right|_{y=R}.
\end{equation}
This means, that $\left. h_{\mu\nu}\right|_{y=R}$ satisfies
transverse-traceless gauge conditon
\begin{eqnarray}
\left. \partial^\nu h_{\mu\nu}\right|_{y=R}=0, \\ \nonumber \left.
h_{\nu}^{\nu}\right|_{y=R}=0,
\end{eqnarray}
which is quite natural, because there is no matter on brane~2.

Thus, the Newtonian limit in this case has the following form \\ 1)
$1<<kr$
\begin{equation}\label{NL2}
V\approx
-\frac{4}{3}G_{1}e^{-kR}\frac{M}{r}\left(1+\sqrt{\frac{2}{\pi}}e^{-1}
\frac{\cos(kr)}{kr}\right),\,
\end{equation}
2) $kr<<1$
\begin{eqnarray}
V\approx
-G_{1}e^{-kR}Mk\frac{8}{3\pi}\left(1+\frac{3\pi}{4\sqrt{2}}-0.2\sqrt{\frac{\pi}{2}}+
0.099\sqrt{\frac{\pi}{2}}\frac{1}{6}(kr)^2\right)\approx \\
\nonumber \approx -G_{1}e^{-kR}Mk\left(2.05+0.02(kr)^2\right).\,
\end{eqnarray}

Below the results for the last two cases are presented. All
calculations were made similarly to the presented above.

I) The mass and the observer are located on brane~2:
\begin{eqnarray}\label{NL3}
V\approx
-G_{2}\frac{M}{r}\left(1-e^{2kR}+\frac{8e^{2kR}}{3\pi}\,si(kr)+
\frac{2e^{2kR}}{3\pi}\left[\frac{\cos(kr)}{kr}+\frac{\sin(kr)}{(kr)^2}\right]\right),
\end{eqnarray}
where $G_2=\frac{\hat G k}{e^{2kR}-1}$ is the effective
gravitational constant on brane 2.

II) The mass is located on brane~2 and the observer is located on
brane~1 ("shadow"\ matter also): \\ 1) $e^{kR}<<kr$
\begin{eqnarray}\label{NL4}
V\approx
-\frac{4}{3}e^{kR}G_{2}\frac{M}{r}\left(1+\sqrt{\frac{2}{\pi}}e^{kR}e^{-1}
\frac{\cos(kre^{-kR})}{kr}\right),\,
\end{eqnarray}
2) $kr<<e^{kR}$
\begin{eqnarray}
V &\approx&
-G_{2}Mk\frac{8}{3\pi}\left(1+\frac{3\pi}{4\sqrt{2}}-0.2\sqrt{\frac{\pi}{2}}+
0.099\,
e^{-2kR}\sqrt{\frac{\pi}{2}}\frac{1}{6}(kr)^2\right)\approx
\\ \nonumber &\approx& -G_{2}Mk\left(2.05+0.02\,e^{-2kR}(kr)^2\right).\,
\end{eqnarray}
We would like to note that in all formulas for the Newtonian
limit, presented above, parameter $M$ denotes the physical mass in
Galilean coordinates on the brane the mass is  located on.

In paper \cite{BKSV} possible values of parameter $k$ were discussed. It
was shown that the value  $k\sim 1 Tev$ is the only one admissible for a
correct physical interpretation of the theory on brane~2. This means, that
for $r\sim 1cm$ the value $kr\sim 10^{17}$ and the terms proportional to
$\sin(kr)$ and $\cos(kr)$ can be dropped as negligible. For example, since
$si(t\to\infty)\to \frac{\pi}{2}$, equation (\ref{NL3}) takes the form
\begin{equation}
V\approx -G_{2}\frac{M}{r}\left(1+\frac{1}{3}e^{2kR}\right).
\end{equation}
The same arguments can be applied to other cases of matter and observer
disposition. Since the contribution of massive modes are negligible for
$r>>10^{-17} cm$, only the zero modes constitute the Newtonian limit.
Unfortunately,  in formulas (\ref{NL1}), (\ref{NL2}), (\ref{NL3}) and
(\ref{NL4}) the  contributions of the  massless graviton and of the radion
are "mixed". This problem can be solved with the help of the zero mode
approximation.

\section{Zero mode approximation}

Let us consider equations (\ref{mu-nu}), (\ref{mu-4}) and (\ref{mu-4}) with
the energy-momentum tensor of the form $T_{\mu\nu} =
t_{\mu\nu}(x)\delta(y)$ (in the unitary gauge). As described above, with
the help of (\ref{substitution1}) these equation (except for the
$\mu\nu$-equation) take the form (\ref{mu4TT}), (\ref{44TT}) and
(\ref{contTT}). Then we can get equations (\ref{eqphi}) and (\ref{equ}) as
it was made above. It is well known that the field $u_{\mu\nu}$ in the
presence of matter is a combination of zero and massive modes
\cite{GarTan}, whose eigenfunctions are orthogonal \cite{BKSV}. In
particular, the zero mode can be represented as
$u^0_{\mu\nu}=e^{2\sigma}\alpha_{\mu\nu}$, where $\alpha_{\mu\nu}$ depends
on $x$ only. It also means that with the help of the residual gauge
transformations $\xi_\mu = e^{2\sigma}\epsilon_\mu(x)$ it is possible to
impose the gauge condition
\begin{eqnarray}\label{gaugeu}
\partial^{\nu}u^m_{\mu\nu}=0, \\ \nonumber
u_{\mu}^{m\mu}=0,
\end{eqnarray}
on the massive modes $u^m_{\mu\nu}$ and the  de Donder gauge condition
on the zero mode
\begin{equation}\label{dedonder2}
\partial^\nu\left(\alpha_{\mu\nu}-\frac{1}{2}\eta_{\mu\nu}
\alpha\right)=0.
\end{equation}
After imposing this gauge, we are still left with residual gauge
transformation (\ref{ostatgauge}). It follows from equations
(\ref{44TT}), (\ref{eqphi}),(\ref{gaugeu}), (\ref{dedonder})  that
\begin{equation}
\Box \alpha=\frac{\hat\kappa c e^{2kR}}{R}t.
\end{equation}

Now let us consider equation (\ref{mu-nu}). Making substitution
(\ref{substitution1}) with condition (\ref{eqphi}) and passing to
the gauge (\ref{gaugeu}), (\ref{dedonder2}) we get
\begin{eqnarray}\label{munuTT}
\frac{1}{2}\Box(\alpha_{\mu\nu}-\frac{1}{2}\eta_{\mu\nu}\alpha)
 + \frac{1}{2}e^{-2\sigma}\Box
u^m_{\mu\nu}+\frac{1}{2}\partial_4
\partial_4 u^m_{\mu\nu}-2k^2u^m_{\mu\nu}-\partial_4
\partial_4\sigma u^m_{\mu\nu}= \\ \nonumber
=-\frac{\hat\kappa}{2}t_{\mu\nu}\delta(y)+\frac{\hat\kappa k
e^{2kR}}{6(e^{2kR}-1)}\left(\frac{\partial_\mu\partial_\nu}{\Box}-
\eta_{\mu\nu}\right)t -\frac{\hat \kappa}{6}
\left(\frac{\partial_\mu\partial_\nu}{\Box}-\eta_{\mu\nu}\right)
t\,\delta(y).
\end{eqnarray}
Since we are going to calculate the  Newtonian limit and light deflection
in the  zero mode approximation, we have to find an equation for the
field $\alpha_{\mu\nu}$. If we multiply equation (\ref{munuTT}) by
$e^{2\sigma}$, integrate it over $y$ and take into account the  orthonormality
condition for the wave functions of the modes, we get
\begin{equation}
\Box (\alpha_{\mu\nu}-\frac{1}{2}\eta_{\mu\nu}\alpha)
=-\frac{\hat\kappa k}{(1-e^{-2kR})}t_{\mu\nu}.
\end{equation}
The equation for massive modes takes the form
\begin{eqnarray}
&&\frac{1}{2}e^{-2\sigma}\Box u^m_{\mu\nu}+\frac{1}{2}\partial_4
\partial_4 u^m_{\mu\nu}-2k^2u^m_{\mu\nu}-\partial_4
\partial_4\sigma u^m_{\mu\nu}=-\frac{\hat\kappa}{2}t_{\mu\nu}\delta(y)+
 \\ \nonumber
&&+ \frac{\hat\kappa k e^{2kR} }{2(e^{2kR}-1)}t_{\mu\nu} +\frac{\hat\kappa k
e^{2kR}}{6(e^{2kR}-1)}\left(\frac{\partial_\mu\partial_\nu}{\Box}-
\eta_{\mu\nu}\right)t-\frac{\hat \kappa
}{6}\left(\frac{\partial_\mu\partial_\nu}{\Box}-
\eta_{\mu\nu}\right)t\,\delta(y).
\end{eqnarray}

We note that equation (\ref{munuTT}) has been solved exactly in Section 2.
But it turned out to be impossible to evaluate analytically all the arising
integrals even for a simple form of $t_{\mu\nu}$ (for example, for a static
point mass).

When  matter is  on brane~2 (at $y=R$), all the reasonings are the
same, as presented above. The substitution has the form
(\ref{subsT2-1}). The gauge conditions are the same as in case of
matter on brane~1 too. The equations of motion for the fields
$\alpha_{\mu\nu}$, $\phi$ in the presence of matter on brane~2 can
be derived analogously and read
\begin{eqnarray}
\frac{1}{2}\Box(\alpha_{\mu\nu}-\frac{1}{2}\eta_{\mu\nu}\alpha) &+&
\frac{1}{2}e^{-2\sigma}\Box u^m_{\mu\nu}+\frac{1}{2}\partial_4
\partial_4 u^m_{\mu\nu}-2k^2u^m_{\mu\nu}-\partial_4
\partial_4\sigma u^m_{\mu\nu}= \\ \nonumber
=-\frac{\hat\kappa}{2}t_{\mu\nu}\delta(y-R)&+&\frac{\hat\kappa k
}{6(e^{2kR}-1)}\left(\frac{\partial_\mu\partial_\nu}{\Box}-\eta_{\mu\nu}\right)
t^{\rho}_{\rho}- \\ \nonumber
 &-& \frac{\hat \kappa}{6}\left(\frac{\partial_\mu\partial_\nu}
{\Box}-\eta_{\mu\nu}\right)t^{\rho}_{\rho}\,\delta(y-R),\\
\Box (\alpha_{\mu\nu}-\frac{1}{2}\eta_{\mu\nu}\alpha)
&=&-\frac{\hat\kappa k}{(e^{2kR}-1)}t_{\mu\nu},\\
\Box \phi=\frac{\hat\kappa}{6R}t_\mu^\mu.&&
\end{eqnarray}
We would like to note once again that these equations are written in the
coordinates $\{x^{\mu}\}$, which are {\it Galilean on brane~1} and are
inappropriate for studying physical effects on brane~2.

Now we are ready to find the  Newtonian limit and light deflection
by a point-like static mass. Let us first make it for brane~1. The
substitution (\ref{substitution1}) in the zero mode approximation
is
\begin{eqnarray}\label{subst0}
h^{0}_{\mu\nu}=e^{2\sigma}\alpha_{\mu\nu}-\frac{kR}{e^{2kR}-1}\eta_{\mu\nu}\phi+
\frac{R}{2k} \partial_\mu\partial_\nu\phi+\frac{1}{4k^2} \partial_\mu\partial_\nu\phi.
\end{eqnarray}
The equation for $h_{\mu\nu}$ on brane~1 in the zero mode
approximation looks like
\begin{eqnarray}\label{h0T}
\Box h^0_{\mu\nu}&=&-\frac{\hat \kappa k
e^{2kR}}{e^{2kR}-1}\left(t_{\mu\nu}-\frac{1}{2}\eta_{\mu\nu}t^{\rho}_{\rho}\right)-
\\ \nonumber -
\frac{\hat \kappa k
}{6(e^{2kR}-1)}\eta_{\mu\nu}t &+& \frac{\hat
\kappa}{12k^2
R}\left(kR+\frac{1}{2}\right)\partial_{\mu}\partial_{\nu}t.
\end{eqnarray}

Let us consider a static point mass with energy-momentum tensor
$t_{00}=M\delta(\vec x), t_{0k}=t_{ij}=0$, $M$ denoting the
inertial mass again. Then we can find a solution of equation
(\ref{h0T}) for $h_{00}$-component
\begin{equation}
h_{00}=\frac{\hat\kappa k
(e^{2kR}+\frac{1}{3})}{8\pi(e^{2kR}-1)}\frac{M}{r},
\end{equation}
where $r=\sqrt{{\vec x}^2}$.
Thus,  we get
\begin{equation}\label{V1}
V=-\hat G k\frac{ (e^{2kR}+\frac{1}{3})}{(e^{2kR}-1)}\frac{M}{r}=
-G_{1}\left(1+\frac{1}{3}e^{-2kR}\right)\frac{M}{r}.
\end{equation}

Now let us show that the Randall-Sundrum model in the zero mode
approximation is equivalent to the linearized Brans-Dicke theory
(apparently, it was first  noted in \cite{GarTan}). Appropriate
formulas for the light deflection in the Brans-Dicke theory are
well known and can be found, for example, in \cite{Weinberg}.

The equation for the fluctuations of metric in the linearized
Brans-Dicke theory  looks like
\begin{eqnarray}\label{BDeq}
\Box \left(h_{\mu\nu}-\frac{1}{2}\eta_{\mu\nu}h\right)=-16\pi G
\left(t_{\mu\nu}-\frac{1}{2\omega+3}\left(
\eta_{\mu\nu}-\frac{\partial_\mu\partial_\nu}{\Box}\right)t\right),
\end{eqnarray}
where $\omega$ is the BD-parameter. It is easy to see that
$h_{\mu\nu}$ satisfies the de Donder gauge condition
$\partial^{\nu}\left(h_{\mu\nu}-\frac{1}{2}\eta_{\mu\nu}h\right)=0$.
The light deflection angle is given by~\cite{Weinberg}
\begin{equation}\label{angle}
\Delta\varphi\approx\frac{4MG}{r_{0}}\left(\frac{2\omega+3}{2\omega+4}\right),
\end{equation}
with $r_{0}$ being the impact parameter and  $M$ being the mass of the
static point-like source.

Now let us examine the four-dimensional effective theory on
brane~1. It is easy to see that $h^{0}_{\mu\nu}$ (\ref{subst0})
does not satisfy the de Donder gauge condition. We can impose it
with the help of the gauge functions (\ref{gfunc}).

In this gauge equation (\ref{h0T}) takes the form
\begin{eqnarray}\label{h0TDD}
\Box \left(h_{\mu\nu}-\frac{1}{2}\eta_{\mu\nu}h\right)=-16\pi
G_{1}\left(t_{\mu\nu}-\frac{e^{-2kR}}{3}\left(\eta_{\mu\nu}-\frac{
\partial_\mu\partial_\nu}{\Box}\right)t\right).
\end{eqnarray}
Here and in what follows we drop the superscript $0$ of the zero
mode approximation. By comparing (\ref{h0TDD}) with (\ref{BDeq}),
we get $\omega = \frac{3}{2}\left(e^{2kR}-1\right) $. Substituting
this into (\ref{angle}), we find the angle of light deflection  by
a static  point-like source  on brane~1
\begin{equation}\label{ang1}
\Delta\varphi\approx\frac{4MG_{1}}{r_{0}}\left(\frac{1}{1+\frac{1}{3}e^{-2kR}}\right).
\end{equation}
The second term in the denominator of formula (\ref{ang1}) and the second
term in formula (\ref{V1}) correspond to the contribution of the
radion. One can see that the contribution of the radion is $e^{2kR}$
times smaller than  the contribution of the massless graviton.

Direct calculations lead us to the following results for the
remaining cases. All the reasonings concerning  Galilean coordinates are
the same as in section~2.

1) We live on brane~1, the mass is located on brane~2 ("shadow"\
matter)
\begin{eqnarray}\label{see}
\Box \left(h_{\mu\nu}-\frac{1}{2}\eta_{\mu\nu}h\right)=-16\pi
G_{1}e^{-kR}\left(t_{\mu\nu}-\frac{1}{3}\left(\eta_{\mu\nu}-\frac{
\partial_\mu\partial_\nu}{\Box}\right)t\right).
\end{eqnarray}
The light deflection angle is
\begin{equation}\label{ang2}
\Delta\varphi\approx\frac{3MG_{1}e^{-kR}}{r_{0}}=4MG_{1}e^{-kR}\left(\frac{1}
{1+\frac{1}{3}}\right)\frac{1}{r_{0}}.
\end{equation}
Newton's Law looks like
\begin{equation}\label{V2}
V=-\frac{4}{3}G_{1}e^{-kR}\frac{M}{r}=-\left(1+\frac{1}{3}\right)G_{1}e^{-kR}\frac{M}{r}.
\end{equation}
The second term in the denominator of formula (\ref{ang2}) and the
second term in the brackets in formula (\ref{V2}) correspond to
the contribution of the radion. One can see that the contribution
of the radion is of the same order, as the contribution of the
massless graviton, but the effect of the "shadow"\ matter is
$e^{kR}$ times smaller, than the effect of the ordinary matter
(compare with (\ref{ang1}) and (\ref{V1})). It means that the
effective theory on brane~1 is phenomenologically acceptable, but
not interesting (see \cite{BKSV}).

2) We live on brane~2, and  the  mass is located on the same brane
\begin{eqnarray}
\Box \left(h_{\mu\nu}-\frac{1}{2}\eta_{\mu\nu}h\right)=-16\pi
G_{2}\left(t_{\mu\nu}-\frac{e^{2kR}}{3}\left(\eta_{\mu\nu}-\frac{
\partial_\mu\partial_\nu}{\Box}\right)t\right)
\end{eqnarray}
with $G_{2}=\hat G k\frac{1}{e^{2kR}-1}$.

The light deflection angle  is
\begin{equation}\label{ang3}
\Delta\varphi\approx\frac{4MG_{2}}{r_{0}}\left(\frac{1}{1+\frac{e^{2kR}}{3}}\right).
\end{equation}
Newton's Law looks like
\begin{equation}\label{V3}
V=-G_{2}\left(1+\frac{e^{2kR}}{3}\right)\frac{M}{r}.
\end{equation}
The second term in the denominator of formula (\ref{ang3}) and the
second term in the brackets in formula (\ref{V3}) correspond to the contribution
of the radion. Now the  contribution of the radion is $e^{2kR}$
times stronger than the contribution of the massless graviton. It
means that, in the case of the massless radion, scalar gravity is
realized on brane~2. However, if one assumes some mechanism for
generating the radion mass, for example, the Goldberger-Wise
mechanism \cite{wise}, gravity in the zero mode approximation
becomes tensor and the hierarchy problem solves.

3) We live on brane~2, the mass is located on brane~1 ("shadow"\
matter also)
\begin{eqnarray}
\Box \left(h_{\mu\nu}-\frac{1}{2}\eta_{\mu\nu}h\right)=-16\pi
G_{2}e^{kR}\left(t_{\mu\nu}-\frac{1}{3}\left(\eta_{\mu\nu}-\frac{
\partial_\mu\partial_\nu}{\Box}\right)t\right).
\end{eqnarray}
The light deflection angle is
\begin{equation}\label{ang4}
\Delta\varphi\approx\frac{4MG_{2}e^{kR}}{r_{0}}\left(\frac{1}{1+\frac{1}{3}}\right)
=\frac{3MG_{1}e^{-kR}}{r_{0}}.
\end{equation}
Newton's Law looks like
\begin{equation}\label{V4}
V=-\left(1+\frac{1}{3}\right)G_{2}e^{kR}\frac{M}{r}=-\frac{4}{3}G_{1}e^{-kR}\frac{M}{r}.
\end{equation}
As  before, the second term in the denominator of formula
(\ref{ang4}) and the second term in the brackets in formula
(\ref{V4}) correspond to the contribution of the  radion. We can
see that the contribution of the radion is of the same order, as
contribution of the massless graviton, but the contribution of the
"shadow"\ matter to  the Newtonian limit is $e^{kR}$ times
smaller, than the contribution of the ordinary matter, because of
the interaction with the massless radion (compare with
(\ref{V3})). But if the radion is massive (otherwise we have
scalar gravity in "our"\ world on brane~2), the contribution of
the "shadow"\ matter to the Newtonian limit is $e^{kR}$ times
stronger, than the contribution of the ordinary matter. It could
lead  to some observable effects, resulting from the  distribution
of matter in the "mirror"\ world (brane~1).

One can shaw that equations (\ref{V1}), (\ref{V2}), (\ref{V3}) and
(\ref{V4}) coincide with the zero-mode parts of equations (\ref{NL1}),
(\ref{NL4}), (\ref{NL3}) and (\ref{NL2}) respectively.

We would like to note that in paper \cite{GarTan} the effects of
the "shadow"\ matter were also considered. In this paper the
deflection angles with the same impact parameters and by the same
Newtonian masses were compared (Newtonian mass is the coefficient
in front of the $\frac{1}{r}$ term in Newton's Law). We have
compared the effects of the "shadow"\ and the ordinary matter, as
it was made in \cite{GarTan}, and we have got the following
results: 44\% instead of 25\% in \cite{GarTan} for brane~1 and the
difference of the order of $e^{4kR}$ instead of 25\% in
\cite{GarTan} for brane~2. This discrepancy appears because we use
the  Galilean coordinates on brane~2. In the case of the massive
radion there is no difference in the zero mode approximation at
all for both branes.

\bigskip
{ \large \bf Acknowledgments}
\medskip

The authors are grateful to E.E.~Boos, Yu.V.~Grats, Yu.A.~Kubyshin,
A.A.~Rossikhin and V.A.~Rubakov for valuable discussions. The work  was
supported by the grant 990588 of the programme "Universities of Russia"\,.
I.V. was also supported in part by the programme SCOPES (Scientific
co-operation between Eastern Europe and Switzerland) of the Swiss National
Science Foundation (project No. 7SUPJ062239) and financed by the Swiss
Federal Department of Foreign Affairs.

\end{document}